\documentclass[12pt]{iopart}
\usepackage{graphicx}

\usepackage{color}

\begin{document}

\title[A simplified M{\o}lmer--S{\o}rensen gate for the trapped ion quantum computer]{A simplified M{\o}lmer--S{\o}rensen gate for the trapped ion quantum computer}

\author{Hiroo Azuma}

\address{Global Research Center for Quantum Information Science,
National Institute of Informatics,
2-1-2 Hitotsubashi, Chiyoda-ku, Tokyo 101-8430, Japan}
\ead{hiroo.azuma@m3.dion.ne.jp, zuma@nii.ac.jp}
\vspace{10pt}
\begin{indented}
\item[]July 2023
\end{indented}

\begin{abstract}
We discuss how to simplify the M{\o}lmer--S{\o}rensen (MS) gate
which is used for the trapped ion quantum computer.
The original MS gate is implemented by illuminating two ions with bichromatic coherent light fields separately at the same time.
In this paper, we propose a method for transforming a separable state of two ions into one of the Bell states
by illuminating the two ions with monochromatic coherent light fields individually
and this point is the advantage of our scheme over the original MS gate.
The length of the execution time of our proposed gate is comparable to that of the original MS gate,
however,
numerical calculations show that our proposed gate is weakly sensitive to thermal fluctuations of the phonons.
By giving another example of a simple two-ion gate that can generate entanglement but is strongly vulnerable to thermal fluctuations,
we show that our simplified MS gate is more marked than usual.
\end{abstract}

%
\noindent{\it Keywords}:
ion trap quantum computing,
Bell states,
entanglement,
warm ion,
M{\o}lmer--S{\o}rensen gate

\submitto{\PS}
%
%
%
\section{Introduction}
Since Preskill put forward the concept of noisy intermediate-scale quantum (NISQ) technology,
many researchers have devoted themselves to the development of quantum computers
composed of about $50$ to $100$ qubits \cite{Preskill2018}.
The most well-known work that heralded the beginning of the NISQ era was the demonstration of quantum supremacy
by Arute {\it et al}. using a 53-qubit quantum computer with the Josephson junctions \cite{Arute2019}.
Arute {\it et al}. performed a task of sampling the output of a pseudo-random circuit with the programmable quantum computer
made up of 53 functional qubits and obtained a million samples within $200$ seconds.
They asserted that it took ten thousand years to finish this task with the most modern supercomputer.
Although superconducting quantum computing attracts the attention of researchers
and the result of Arute {\it et al}. is considered to be a notable milestone,
there are alternative methods for constructing quantum processors \cite{Knill2001,Raussendorf2003,Khazali2020}.
The most typical approach of these alternatives is ion trap quantum computing.
The practical realization of a fully controlled 20-qubit processor for trapped ion quantum computation was reported \cite{Friis2018},
and we can recognize it as the NISQ computer.

The first theoretical proposal of the trapped ion quantum computation was provided by Cirac and Zoller in 1995 \cite{Cirac1995}.
Since this proposition,
aggressive research activities have been conducted
and experimental technologies for the actual implementation of ion trap quantum computing were extremely refined in recent years.
For example,
implementation of a full quantum algorithm was accomplished with the trapped ion system in 2003 \cite{Gulde2003}.
How to scale up a small quantum register composed of trapped ions to large numbers of qubits was discussed \cite{Kielpinski2002}.
A many-body dynamical phase transition with a $53$-qubit quantum simulator was demonstrated using spins of trapped ions \cite{Zhang2017}.

The trapped ion quantum computer has the following salient features.
It was reported that a coherence time of over ten minutes was observed for a single qubit in a trapped ion system \cite{Wang2017}.
A qubit was constructed from a single trapped calcium ion and its readout fidelity attained $99.77$\% \cite{Myerson2008}.
By using the hyperfine states in the ground 4S${}_{1/2}$ levels of the ${}^{43}$Ca${}^{+}$ ion,
qubit state preparation and single-shot readout fidelity reached $99.93$\% \cite{Harty2014}.
A single-qubit gate fidelity of $99.9999$\% was also shown \cite{Harty2014}.
The ion trap quantum computation is suitable for generating entanglement between two qubits, as well.
Experimentally,
an error rate for generation of the Bell states was estimated
at $8\times 10^{-4}$ \cite{Gaebler2016},
and fidelity of the two-qubit gate attained $99.9$\% \cite{Ballance2016}.
A protocol for simultaneously entangling arbitrary pairs of qubits on an ion-trap quantum computer was proposed theoretically \cite{Grzesiak2020}.
Thus, this protocol realizes multiple two-qubit gates.
Comprehensive reviews of ion trap quantum computing were provided
\cite{Nielsen2002,Steane1997,Wineland1998,Haffner2008,Rieffel2011,Zygelman2018,Bruzewicz2019,LaPierre2021}.

Although Cirac and Zoller's original protocol is excellent,
it is vulnerable to thermal noise caused by fluctuations of phonons at finite temperatures.
(The phonon is the center-of-mass vibrational mode of the ion chain in the trapped ion quantum computer.)
An attempt to resolve this problem was made in Refs.~\cite{Molmer1999,Sorensen1999}
by M{\o}lmer and S{\o}rensen.
The implementation of their gate was illuminating two ions with a bichromatic laser light individually and simultaneously.
In this scenario,
if we calculate a perturbation theory of interaction among two ions and the phonon mode,
the first-order energy shift disappears and the second-order correction becomes essential.
When we sum all second-order terms to which intermediate states contribute,
miracle interference occurs and the total energy shift does not rely on the number of phonons at all.
Thus, even if the phonons obey the thermal distribution,
the M{\o}lmer--S{\o}rensen (MS) gate applied to the two ions does not suffer from thermal noise at all.

In this paper, we propose a simplified MS gate.
We can realize this proposed gate by illuminating two ions A and B with monochromatic beams of the laser fields individually at the same time.
The MS gate requires two spatially separated bichromatic laser beams.
In contrast, to accomplish our proposed gate, we need two monochromatic laser beams, and this point is the advantage over the original MS gate.
Here, we describe the period of the time evolution of the wave function for the ions A and B
caused by the proposed gate as $T_{0}(n)$, where $n$ represents the number of phonons in the initial state.
If we let the execution time of the gate be equal to $T_{0}(n)/2$,
it realizes the following transformation:
\begin{eqnarray}
|g\rangle_{\mbox{\scriptsize A}}|g\rangle_{\mbox{\scriptsize B}}|n\rangle_{\mbox{\scriptsize P}}
&\to&
|e\rangle_{\mbox{\scriptsize A}}|e\rangle_{\mbox{\scriptsize B}}|n\rangle_{\mbox{\scriptsize P}}, \nonumber \\
|e\rangle_{\mbox{\scriptsize A}}|e\rangle_{\mbox{\scriptsize B}}|n\rangle_{\mbox{\scriptsize P}}
&\to&
|g\rangle_{\mbox{\scriptsize A}}|g\rangle_{\mbox{\scriptsize B}}|n\rangle_{\mbox{\scriptsize P}}, \nonumber \\
|g\rangle_{\mbox{\scriptsize A}}|e\rangle_{\mbox{\scriptsize B}}|n\rangle_{\mbox{\scriptsize P}}
&\to&
e^{i\varphi_{1}(n)}
|g\rangle_{\mbox{\scriptsize A}}|e\rangle_{\mbox{\scriptsize B}}|n\rangle_{\mbox{\scriptsize P}}, \nonumber \\
|e\rangle_{\mbox{\scriptsize A}}|g\rangle_{\mbox{\scriptsize B}}|n\rangle_{\mbox{\scriptsize P}}
&\to&
e^{i\varphi_{2}(n)}
|e\rangle_{\mbox{\scriptsize A}}|g\rangle_{\mbox{\scriptsize B}}|n\rangle_{\mbox{\scriptsize P}},
\end{eqnarray}
where $|g\rangle$ and $|e\rangle$ are ground and excited states of the ions A and B respectively,
$|n\rangle_{\mbox{\scriptsize P}}$ is a number state of phonons,
and $\varphi_{1}(n)$ and $\varphi_{2}(n)$ are specified functions of $n$ for the phases.
We can show numerically that
$T_{0}(n)\simeq T_{0}(0)+c n$
and $c$ is a small positive constant such as $c/T_{0}(0)\ll 1$. 
Moreover, if we let the execution time be equal to $T_{0}(n)/4$,
the proposed gate generates entanglement as
\begin{equation}
|g\rangle_{\mbox{\scriptsize A}}
|g\rangle_{\mbox{\scriptsize B}}
\to
|\Psi\rangle_{\mbox{\scriptsize AB}}
=
\frac{e^{-i\pi/4}}{\sqrt{2}}
(
|g\rangle_{\mbox{\scriptsize A}}
|g\rangle_{\mbox{\scriptsize B}}
+
i
|e\rangle_{\mbox{\scriptsize A}}
|e\rangle_{\mbox{\scriptsize B}}
).
\label{generation-maximally-entangled-state-0}
\end{equation}
In this way, by adjusting the execution time,
we can obtain the maximally entangled state from the initial separable state.
Because an arbitrary unitary operation for a single ion can be performed by Cirac and Zoller's single-ion gate
and it works much faster than conditional gates,
we can construct all four Bell states from $|\Psi\rangle_{\mbox{\scriptsize AB}}$ in Eq.~(\ref{generation-maximally-entangled-state-0}) efficiently.
However, numerical calculations show that our proposed gate is not robust under thermal fluctuations of the phonons.
In detail, if we draw a curve of the fidelity $F$ of our proposed gate as a function of the average number of the thermal phonons
$n_{\mbox{\scriptsize th}}$,
it does not satisfy $dF/dn_{\mbox{\scriptsize th}}=0$.

In \cite{Figgatt2019}, parallel entangling operations for ion trap quantum computation are experimentally demonstrated.
In \cite{Pogorelov2021}, how to build a compact ion-trap quantum computing system is discussed and its operations are confirmed experimentally.
The objectives of these works are similar to our method.

In this paper, we do not discuss how to mitigate the motion heating that is caused by the electric field noise.
If we build a scalable trapped-ion quantum computer,
we must make the entire device smaller and the distance between neighboring trapped ions becomes shorter.
This situation raises the ion motion heating rate \cite{Bruzewicz2015}.
The simplest method to reduce the motion heating rate is cooling the trap electrodes to the cryogenic temperature,
for example, approximately $4$ K \cite{Chiaverini2014}.
Other methods to decrease the rate were mentioned in \cite{Bruzewicz2015}.
We can apply those techniques to our proposed gate.

The light-shift (LS) gate \cite{Leibfried2003} is another common method for implementation of two-qubit gates of the trapped ion quantum computer which is comparable to the MS gate.
Applying the displacement operators of the phonons in the `stretch' mode,
where the displacements of the two ions are equal but in opposite directions, sequentially,
this method gives a geometric phase to the entire state of the two ions and phonons.
Because the geometric phase is equal to the enclosed area of the phase space,
the LS gate is robust and insensitive to the starting state of the phonons,
so that we do not need ground-state cooling.
In \cite{Sawyer2021,Clark2021}, a variant of the LS gate where the qubit levels are separated by an optical frequency was proposed and experimentally demonstrated.
The LS gate, the MS gate, and our proposed gate share points in common,
for example,
the number of phonons hardly affects the operations of the LS and MS gates and they do not require ground-state cooling,
and we try to analyze our proposed gate along the same lines.
However, our method does not utilize the geometrical phase; thus, it might be vulnerable to detuning of frequencies of laser fields.

In \cite{Sorensen2000}, the preparation of entangled states of ions by illumination with bichromatic light for weak-field and strong-field regimes was considered,
and the fidelity of generating maximally entangled states of an arbitrary number of ions with nonideal conditions was calculated.
Specifically, in \cite{Sorensen2000}, not only the center-of-mass motion in the ion trap but also the spectator vibrational modes were assumed.
Moreover, effects caused by heating of the vibrational motion were estimated.
By contrast, this paper does not examine cases where the trapped ions suffer from errors that are inevitable for real experiments except for the thermal noise of the phonons.
This is a challenge for the future.

This paper is organized as follows.
In Sec.~\ref{section-review-MS-gate}, we give a brief review of the MS gate.
In Sec.~\ref{section-simplified-MS-gate}, we explain the implementation of the proposed gate and examine its dynamics numerically.
In Sec.~\ref{section-mathematical-analyses}, we mathematically analyze the time evolution of the proposed gate with an initial state whose number of phonons is equal to zero.
In Sec.~\ref{section-example-two-qubit-gate}, we give an example of a two-qubit gate that is strongly vulnerable to thermal noise.
In Sec.~\ref{section-conclusions-discussions}, we provide conclusions and discussions.
In \ref{appendix-time-evolution-initial-states}, we compute the time evolution of the proposed gate for the initial states
that are not the ground state of the atoms.
In \ref{appendix-first-order-differential-equation}, we provide an explicit form of the Schr{\"{o}}dinger equation
derived from the second-order perturbation theory.
In \ref{appendix-initial-non-zero-phonon-states}, we mathematically estimate the periods of the proposed gate for which the initial state contains one or more phonons.
In \ref{appendix-operation-times-Cirac-Zoller-gate}, we show the estimation of the operation time for Cirac and Zoller's ion--phonon gate.

\section{\label{section-review-MS-gate}A brief review of the MS gate}
The Hamiltonian of the interaction between an ion and phonons for the trapped ion quantum computer is given by
\cite{Cirac1995,Nielsen2002,Steane1997,Wineland1998}
\begin{eqnarray}
H
&=&
\hbar\Omega\sigma_{+}
\exp
\{
i\eta[a\exp(-i\nu t)+a^{\dagger}\exp(i\nu t)]
-i\delta t+i\phi
\} \nonumber \\
&&
+
\hbar\Omega\sigma_{-}
\exp
\{
-i\eta[a\exp(-i\nu t)+a^{\dagger}\exp(i\nu t)]
+i\delta t-i\phi
\},
\label{ion-trap-Hamiltonian-basic-0}
\end{eqnarray}
where $\Omega$ is the Rabi frequency of a beam of the laser field with which the ion is illuminated,
$\sigma_{+}$ and $\sigma_{-}$ are the raising and lowering operators of the ion respectively,
$\eta$ is the Lamb--Dicke parameter with $0\leq\eta\ll 1$,
$a$ and $a^{\dagger}$ are the annihilation and creation operators of the phonons respectively,
$\nu$ is the angular frequency of the phonons,
$\delta=\omega-\omega_{eg}$,
$\omega$ is the angular frequency of the laser field injected into the ion,
$\hbar\omega_{eg}$ is a difference of energy levels between excited and ground states of the ion,
and $\phi$ is a constant phase of the injected laser light.
From now on, for the sake of simplicity, we put $\hbar=1$.

To implement the MS gate,
we illuminate the ion A with a bichromatic beam of laser fields whose angular frequencies are equal to
$\omega_{\mbox{\scriptsize A}1}$
and
$\omega_{\mbox{\scriptsize A}2}$
and the ion B with a bichromatic beam of laser fields whose angular frequencies are equal to
$\omega_{\mbox{\scriptsize B}1}$
and
$\omega_{\mbox{\scriptsize B}2}$,
where
\begin{eqnarray}
&&
\omega_{\mbox{\scriptsize A}1}
=
\omega_{\mbox{\scriptsize B}1}
=
\omega_{eg}-\tilde{\delta}, \nonumber \\
&&
\omega_{\mbox{\scriptsize A}2}
=
\omega_{\mbox{\scriptsize B}2}
=
\omega_{eg}+\tilde{\delta},
\label{MS-gate-omega-AB-12-0}
\end{eqnarray}
\begin{equation}
\tilde{\delta}
=
\nu+\Delta\nu,
\label{MS-gate-delta-0}
\end{equation}
and we assume
$|\Delta\nu/\nu|\ll 1$.
Moreover, we assume that the constant phases and the Lamb--Dicke parameters of the laser beams
are equal to each other, respectively, as
\begin{eqnarray}
&&
\phi_{\mbox{\scriptsize A}1}
=
\phi_{\mbox{\scriptsize A}2}
=
\phi_{\mbox{\scriptsize B}1}
=
\phi_{\mbox{\scriptsize B}2}
=0, \nonumber \\
&&
\eta_{\mbox{\scriptsize A}1}
=
\eta_{\mbox{\scriptsize A}2}
=
\eta_{\mbox{\scriptsize B}1}
=
\eta_{\mbox{\scriptsize B}2}
=\eta.
\label{MS-gate-phi-eta-0}
\end{eqnarray}

Preparing an initial state
$|g\rangle_{\mbox{\scriptsize A}}
|g\rangle_{\mbox{\scriptsize B}}
|n\rangle_{\mbox{\scriptsize P}}$
and injecting the two beams into the two ions as the implementation of the MS gate,
we obtain the maximally entangled state of the ions A and B at time $t=T_{\mbox{\scriptsize MS}}/4$
where
\begin{equation}
T_{\mbox{\scriptsize MS}}
=
\frac{\pi\Delta\nu}{2\eta^{2}\Omega^{2}}.
\label{definition-T-MS-0}
\end{equation}
We draw attention to the fact that the process of this two-qubit gate is not affected by the number of phonons.
Therefore, even if the phonons are in the thermal equilibrium state,
the MS gate works without errors and we can regard it as robust.

\section{\label{section-simplified-MS-gate}A simplified MS gate and its dynamics obtained by numerical calculations}
We propose the following simplified MS gate.
We illuminate the ions A and B with beams of laser fields whose angular frequencies are given by
$\omega_{\mbox{\scriptsize A}}=\omega_{eg}-\nu-\Delta\nu$
and
$\omega_{\mbox{\scriptsize B}}=\omega_{eg}+\nu+\Delta\nu$, respectively.
According to Eq.~(\ref{ion-trap-Hamiltonian-basic-0}),
Hamiltonians of interaction between the ions A and B and the phonons are given by
\begin{eqnarray}
H&=&H_{\mbox{\scriptsize A}}+H_{\mbox{\scriptsize B}}, \nonumber \\
H_{\chi}
&=&
\Omega_{\chi}\sigma_{+\chi}
\exp
\{
i\eta[a\exp(-i\nu t)+a^{\dagger}\exp(i\nu t)]
+
\mbox{sgn}(\chi)
i(\nu+\Delta\nu)t
\} \nonumber \\
&&
+
\Omega_{\chi}\sigma_{-\chi}
\exp
\{
-i\eta[a\exp(-i\nu t)+a^{\dagger}\exp(i\nu t)] \nonumber \\
&&
-
\mbox{sgn}(\chi)
i(\nu+\Delta\nu)t
\} \nonumber \\
&&
\quad
\mbox{for $\chi\in\{\mbox{A},\mbox{B}\}$}, \nonumber \\
\mbox{sgn}(\chi)
&=&
\left\{
\begin{array}{ll}
1 & \mbox{for $\chi=\mbox{A}$}, \\
-1 & \mbox{for $\chi=\mbox{B}$},
\end{array}
\right.
\label{Hamiltonian-0}
\end{eqnarray}
where we assume
$\phi_{\mbox{\scriptsize A}}=\phi_{\mbox{\scriptsize B}}=0$
and
$\eta_{\mbox{\scriptsize A}}=\eta_{\mbox{\scriptsize B}}=\eta$.
From now on, for the sake of simplicity,
we put
$\Omega_{\mbox{\scriptsize A}}=\Omega_{\mbox{\scriptsize B}}=\Omega>0$.
Regarding $\eta$ as a perturbation parameter and considering the zeroth, first, and second-order terms of
$H_{\mbox{\scriptsize A}}$ and $H_{\mbox{\scriptsize B}}$ in $\eta$,
we obtain
\begin{eqnarray}
H_{\chi}
&=&
H_{\chi}^{(0)}
+
H_{\chi}^{(1)}
+
H_{\chi}^{(2)}, \nonumber \\
H_{\chi}^{(0)}
&=&
\Omega\sigma_{+\chi}
e^{\mbox{\scriptsize sgn}(\chi)i(\nu+\Delta\nu)t}
+
\Omega\sigma_{-\chi}
e^{-\mbox{\scriptsize sgn}(\chi)i(\nu+\Delta\nu)t}, \nonumber \\
H_{\chi}^{(1)}
&=&
i\eta
\Omega\sigma_{+\chi}
e^{\mbox{\scriptsize sgn}(\chi)i(\nu+\Delta\nu)t}(a e^{-i\nu t}+a^{\dagger}e^{i\nu t}) \nonumber \\
&&
-
i\eta
\Omega\sigma_{-\chi}
e^{-\mbox{\scriptsize sgn}(\chi)i(\nu+\Delta\nu)t}(a e^{-i\nu t}+a^{\dagger}e^{i\nu t}), \nonumber \\
H_{\chi}^{(2)}
&=&
-\frac{\eta^{2}}{2}
\Omega\sigma_{+\chi}
e^{\mbox{\scriptsize sgn}(\chi)i(\nu+\Delta\nu)t}(a e^{-i\nu t}+a^{\dagger}e^{i\nu t})^{2} \nonumber \\
&&
-
\frac{\eta^{2}}{2}
\Omega\sigma_{-\chi}
e^{-\mbox{\scriptsize sgn}(\chi)i(\nu+\Delta\nu)t}(a e^{-i\nu t}+a^{\dagger}e^{i\nu t})^{2} \nonumber \\
&&
\quad
\mbox{for $\chi\in\{\mbox{A},\mbox{B}\}$}.
\label{Hamiltonian-AB-second-order-1}
\end{eqnarray}

\begin{figure}
\begin{center}
\includegraphics{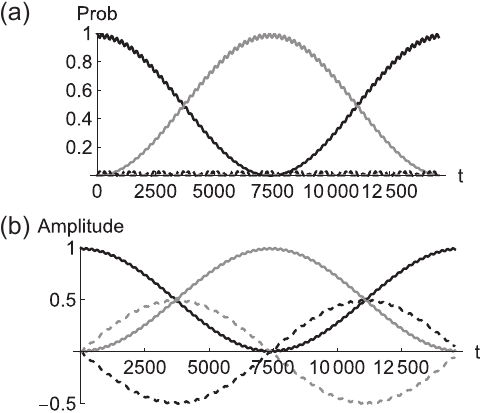}
\end{center}
\caption{
(a) Plots of $\mbox{Prob}(kl)$ for $k,l\in\{g,e\}$ given by Eq.~(\ref{Prob-kl-0}) as functions of time $t$
for the initial state
$|\psi(t=0)\rangle=
|g\rangle_{\mbox{\scriptsize A}}
|g\rangle_{\mbox{\scriptsize B}}
|0\rangle_{\mbox{\scriptsize P}}$.
The solid black, solid grey, dashed black, and dashed grey curves represent graphs of
$\mbox{Prob}(gg)$, $\mbox{Prob}(ee)$, $\mbox{Prob}(ge)$, and $\mbox{Prob}(eg)$, respectively.
Because $\mbox{Prob}(eg)$ is nearly equal to zero $\forall t$,
its curve overlaps the horizontal axis.
We set the physical parameters as
$\Omega=0.1$, $\eta=0.025$, $\nu=5$, and $\Delta\nu=0.025$.
Looking at this figure,
we note that the period of the graphs is equal to
$T_{0}(0)\simeq 14608$
and the state becomes maximally entangled
at $t=T_{0}(0)/4\simeq 3652$.
(b) Plots of $\mbox{Re}(c_{gg})$, $\mbox{Im}(c_{gg})$, $\mbox{Re}(c_{ee})$, and $\mbox{Im}(c_{ee})$
given by Eq.~(\ref{c-gg-c-ee-0})
as functions of time $t$.
The solid black, dashed black, solid grey, and dashed grey curves represent the graphs of
$\mbox{Re}(c_{gg})$, $\mbox{Im}(c_{gg})$, $\mbox{Re}(c_{ee})$, and $\mbox{Im}(c_{ee})$,
respectively.
The physical parameters are given by the same as (a).
}
\label{Figure1}
\end{figure}

Next, we examine how the proposed quantum gate works by numerical calculations.
Considering the Hamiltonian defined in Eq.~(\ref{Hamiltonian-0})
up to the second-order perturbation in $\eta$
as shown in
Eq.~(\ref{Hamiltonian-AB-second-order-1}),
we compute the time evolution of the wave function of the gate with the fourth-order Runge--Kutta method.
Assuming that the initial state is given by
$|\psi(t=0)\rangle
=
|g\rangle_{\mbox{\scriptsize A}}|g\rangle_{\mbox{\scriptsize B}}|0\rangle_{\mbox{\scriptsize P}}$,
we plot the following four probabilities as functions of time $t(\geq 0)$ in Fig.~\ref{Figure1}(a):
\begin{equation}
\mbox{Prob}(kl)
=
\sum_{n}
|{}_{\mbox{\scriptsize A}}\langle k|{}_{\mbox{\scriptsize B}}\langle l|{}_{\mbox{\scriptsize P}}\langle n|\psi(t)\rangle|^{2}
\quad
\mbox{for $k,l\in\{g,e\}$}.
\label{Prob-kl-0}
\end{equation}
Defining amplitudes of $|\psi(t)\rangle$ as
\begin{eqnarray}
c_{gg}
&=&
{}_{\mbox{\scriptsize A}}\langle g|{}_{\mbox{\scriptsize B}}\langle g|{}_{\mbox{\scriptsize P}}\langle 0|\psi(t)\rangle, \nonumber \\
c_{ee}
&=&
{}_{\mbox{\scriptsize A}}\langle e|{}_{\mbox{\scriptsize B}}\langle e|{}_{\mbox{\scriptsize P}}\langle 0|\psi(t)\rangle,
\label{c-gg-c-ee-0}
\end{eqnarray}
we plot
$\mbox{Re}(c_{gg})$, $\mbox{Im}(c_{gg})$, $\mbox{Re}(c_{ee})$, and $\mbox{Im}(c_{ee})$
as functions of time $t(\geq 0)$ in Fig.~\ref{Figure1}(b).

From the results of numerical calculations
for $\Delta\nu>0$ and $\Omega>0$,
we observe the following facts empirically:
Setting
$|\psi(t=0)\rangle
=
|g\rangle_{\mbox{\scriptsize A}}|g\rangle_{\mbox{\scriptsize B}}|n\rangle_{\mbox{\scriptsize P}}$
for $n=0,1,...$ and letting
$T_{0}(n)$ be a period of the time evolution of $|\psi(t)\rangle$,
we obtain an approximate form of $|\psi(t)\rangle$ as 
\begin{eqnarray}
|\psi(t)\rangle
&\simeq&
\frac{1}{2}
[1+\exp(-2\pi i\frac{t}{T_{0}(n)})]
|g\rangle_{\mbox{\scriptsize A}}|g\rangle_{\mbox{\scriptsize B}}|n\rangle_{\mbox{\scriptsize P}} \nonumber \\
&&
+
\frac{1}{2}
[1-\exp(-2\pi i\frac{t}{T_{0}(0)})]
|e\rangle_{\mbox{\scriptsize A}}|e\rangle_{\mbox{\scriptsize B}}|n\rangle_{\mbox{\scriptsize P}}.
\end{eqnarray}
In Sec.~\ref{section-mathematical-analyses},
we show that $T_{0}(n)$ does not change largely as $n$ increases from $0$ to $1,2,...$.
If we set $t=T_{0}(n)/4$, the state is given by
$|\psi(T_{0}(n)/4)\rangle
\simeq
|\Psi\rangle_{\mbox{\scriptsize AB}}|n\rangle_{\mbox{\scriptsize P}}$
where
$|\Psi\rangle_{\mbox{\scriptsize AB}}$
is defined in Eq.~(\ref{generation-maximally-entangled-state-0}),
so that we obtain the maximally entangled state.
We present approximate expressions of the time evolution for the initial states
$|\psi(t=0)\rangle=|e\rangle_{\mbox{\scriptsize A}}|e\rangle_{\mbox{\scriptsize B}}|n\rangle_{\mbox{\scriptsize P}}$,
$|g\rangle_{\mbox{\scriptsize A}}|e\rangle_{\mbox{\scriptsize B}}|n\rangle_{\mbox{\scriptsize P}}$,
and
$|e\rangle_{\mbox{\scriptsize A}}|g\rangle_{\mbox{\scriptsize B}}|n\rangle_{\mbox{\scriptsize P}}$
for $n=0,1,...$
in \ref{appendix-time-evolution-initial-states}.

Next, we consider the time evolution of the state of the proposed gate when the phonons are in thermal equilibrium.
According to the Bose--Einstein statistics,
the probability that the number of phonons is equal to $n$ is given by
\begin{equation}
p_{n}
=
[1-\exp(-\beta\nu)]\exp(-\beta n\nu),
\end{equation}
where $\beta=1/(k_{\mbox{\scriptsize B}}T)$, $k_{\mbox{\scriptsize B}}$ is the Boltzmann constant, and $T$ represents the temperature.
The mean number of the phonons in the thermal equilibrium is given by
\begin{equation}
n_{\mbox{\scriptsize th}}
=
[\exp(\beta\nu)-1]^{-1}.
\end{equation}
Thus, we can regard $n_{\mbox{\scriptsize th}}$ as a variable that characterizes the state in thermal equilibrium instead of the temperature $T$.

\begin{figure}
\begin{center}
\includegraphics{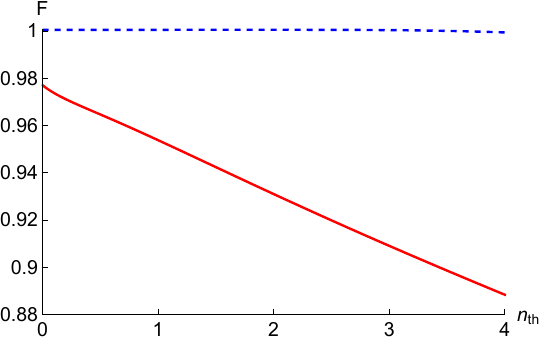}
\end{center}
\caption{
A plot of the fidelity $F$ of our proposed gate given by Eq.~(\ref{fidelity-0}) and that of the original MS gate
as functions of $n_{\mbox{\scriptsize th}}$,
where they are represented by the solid red and dashed blue curves, respectively.
Evaluating the fidelity by numerical calculations,
we assume that the dimension of the Hilbert space of the phonons is equal to $31$ and its basis vectors are given by
$\{|0\rangle_{\mbox{\scriptsize P}},|1\rangle_{\mbox{\scriptsize P}},
...,
|30\rangle_{\mbox{\scriptsize P}}\}$.
For our proposed gate,
the physical parameters are given by the same as Fig.~\ref{Figure1}.
For the original MS gate,
we set parameters as
$\Omega=0.05$, $\eta=0.005$, $\nu=10$, and $\Delta\nu=0.005$ for Eqs.~(\ref{ion-trap-Hamiltonian-basic-0}), (\ref{MS-gate-omega-AB-12-0}), (\ref{MS-gate-delta-0}), and (\ref{MS-gate-phi-eta-0}).
The fidelity of our proposed method does not start from unity for $n_{\mbox{\scriptsize th}}=0$.
}
\label{Figure2}
\end{figure}

Letting the initial state be given by
\begin{equation}
\rho(t=0)
=
\sum_{n=0}^{\infty}
p_{n}
|g\rangle_{\mbox{\scriptsize A}}{}_{\mbox{\scriptsize A}}\langle g|
\otimes
|g\rangle_{\mbox{\scriptsize B}}{}_{\mbox{\scriptsize B}}\langle g|
\otimes
|n\rangle_{\mbox{\scriptsize P}}{}_{\mbox{\scriptsize P}}\langle n|,
\end{equation}
we cause its time evolution until $t=T_{0}(0)/4$ up to the second-order Hamiltonian in $\eta$
that is defined in Eq.~(\ref{Hamiltonian-AB-second-order-1}).
We define the fidelity of the state $\rho(t=T_{0}(0)/4)$ and the maximally entangled state of the ions A and B,
that is to say,
$|\Psi\rangle_{\mbox{\scriptsize AB}}$, as
\begin{equation}
F
=
{}_{\mbox{\scriptsize AB}}\langle\Psi|\mbox{Tr}_{\mbox{\scriptsize P}}
[\rho(t=T_{0}(0)/4)]
|\Psi\rangle_{\mbox{\scriptsize AB}}.
\label{fidelity-0}
\end{equation}
We plot $F$ as a function of $n_{\mbox{\scriptsize th}}$ in Fig.~\ref{Figure2}.
Looking at Fig.~\ref{Figure2}, we note that
$dF/dn_{\mbox{\scriptsize th}}$ cannot be equal to zero.
Thus, we must recognize that our proposed gate is not insensitive to fluctuations of the thermal phonons.
For a comparison of our proposed gate and the original MS gate,
we also plot the fidelity for the original MS gate as a function of $n_{\mbox{\scriptsize th}}$ in Fig.~\ref{Figure2}.

\section{\label{section-mathematical-analyses}Mathematical analyses of the time evolution of the proposed gate}
In this section, we examine the dynamics of the proposed gate with a non-rigorous but mathematical method.
In particular, we derive the period of the time evolution of the proposed gate whose initial state is given by
$|g\rangle_{\mbox{\scriptsize A}}|g\rangle_{\mbox{\scriptsize B}}|0\rangle_{\mbox{\scriptsize P}}$
as a function ${\cal T}_{0}(\Omega,\eta,\nu,\Delta\nu)$.
Moreover, based on the results of numerical calculations, we explain that the function of the period of the proposed gate can hardly depend on the number of phonons
in the initial state.

First, for the sake of simplicity, we consider a situation where the number of the phonons in the initial state is equal to zero, that is,
$|\psi(t=0)\rangle=|g\rangle_{\mbox{\scriptsize A}}|g\rangle_{\mbox{\scriptsize B}}|0\rangle_{\mbox{\scriptsize P}}$.
Moreover, we assume that the gate lies on an eight-dimensional Hilbert space whose orthonormal basis is given by
\begin{eqnarray}
&&
|\bar{0}\rangle
=
|g\rangle_{\mbox{\scriptsize A}}|g\rangle_{\mbox{\scriptsize B}}|0\rangle_{\mbox{\scriptsize P}},
\quad
|\bar{1}\rangle
=
|g\rangle_{\mbox{\scriptsize A}}|g\rangle_{\mbox{\scriptsize B}}|1\rangle_{\mbox{\scriptsize P}}, \nonumber \\
&&
|\bar{2}\rangle
=
|g\rangle_{\mbox{\scriptsize A}}|e\rangle_{\mbox{\scriptsize B}}|0\rangle_{\mbox{\scriptsize P}},
\quad
|\bar{3}\rangle
=
|g\rangle_{\mbox{\scriptsize A}}|e\rangle_{\mbox{\scriptsize B}}|1\rangle_{\mbox{\scriptsize P}}, \nonumber \\
&&
|\bar{4}\rangle
=
|e\rangle_{\mbox{\scriptsize A}}|g\rangle_{\mbox{\scriptsize B}}|0\rangle_{\mbox{\scriptsize P}},
\quad
|\bar{5}\rangle
=
|e\rangle_{\mbox{\scriptsize A}}|g\rangle_{\mbox{\scriptsize B}}|1\rangle_{\mbox{\scriptsize P}}, \nonumber \\
&&
|\bar{6}\rangle
=
|e\rangle_{\mbox{\scriptsize A}}|e\rangle_{\mbox{\scriptsize B}}|0\rangle_{\mbox{\scriptsize P}},
\quad
|\bar{7}\rangle
=
|e\rangle_{\mbox{\scriptsize A}}|e\rangle_{\mbox{\scriptsize B}}|1\rangle_{\mbox{\scriptsize P}}.
\end{eqnarray}
This treatment implies that we use the following $2\times 2$ matrices as an annihilation operator of the phonons for the approximation:
\begin{equation}
\hat{a}
=
\left(
\begin{array}{cc}
0 & 1 \\
0 & 0 \\
\end{array}
\right),
\end{equation}
where the basis is given by $\{|0\rangle_{\mbox{\scriptsize P}},|1\rangle_{\mbox{\scriptsize P}}\}$.
Here, we describe $|\psi(t)\rangle$ as a superposition of $\{|\bar{i}\rangle:i\in\{0,1,...,7\}\}$ as follows:
\begin{equation}
|\psi(t)\rangle
=
\sum_{i=0}^{7}
c_{i}(t)|\bar{i}\rangle.
\end{equation}
Then, from the Schr{\"{o}}dinger equation
$
i(\partial/\partial t)
|\psi(t)\rangle
=
\hat{H}|\psi(t)\rangle
$,
we obtain a system of first-order differential equations,
\begin{equation}
i\dot{c}_{j}(t)
=
\sum_{k=0}^{7}
c_{k}(t)\langle\bar{j}|\hat{H}|\bar{k}\rangle
\quad
\mbox{for $j=0,1,...,7$},
\label{Schrodinger-equation-ckt-0}
\end{equation}
\begin{equation}
c_{0}(0)=1,
\quad
c_{k}(0)=0
\quad
\mbox{for $k=1,2,...,7$}.
\label{Schrodinger-equation-ckt-1}
\end{equation}
The matrix elements
$\langle\bar{j}|\hat{H}|\bar{k}\rangle$
for $j,k\in\{0,1,...,7\}$
are derived with Eq.~(\ref{Hamiltonian-AB-second-order-1}).
We give their explicit forms in \ref{appendix-first-order-differential-equation}.

From numerical calculations carried out in Sec.~\ref{section-simplified-MS-gate},
we obtain the following empirical findings.
Here, we assume
$\nu\gg 1$, $\Omega\ll 1$, $\eta\ll 1$, and $\Delta\nu\ll 1$.
Then,
$c_{5}(t)\simeq 0$ and $c_{6}(t)\simeq 1-c_{0}(t)$ hold.
Now, we replace $-2+\eta^{2}$ with $-2$ in the matrix elements because of $\eta^{2}\ll 1$.
Moreover, according to the rotating-wave approximation,
we make terms that include $e^{\pm 2i\nu t}$ equal to zero.
Then, we can single a closed differential equation of $c_{2}(t)$ out from Eqs.~(\ref{Schrodinger-equation-ckt-0}) and (\ref{Schrodinger-equation-ckt-1})
as
\begin{equation}
i\dot{c}_{2}(t)
=
e^{-i(\nu+\Delta\nu)t}\Omega,
\quad
c_{2}(0)=0.
\end{equation}
Thus, we obtain
\begin{equation}
c_{2}(t)
=
-\frac{\Omega}{\nu+\Delta\nu}
[1-e^{-i(\nu+\Delta\nu)t}].
\end{equation}
Further, from empirical evidence obtained by numerical calculations in Sec.~\ref{section-simplified-MS-gate},
we apply the following approximation to the system of differential equations,
\begin{eqnarray}
x(t)&=&\frac{1}{2}[c_{1}(t)+c_{7}(t)], \nonumber \\
c_{1}(t)-c_{7}(t)&\simeq&0.
\label{x-t-approximation-0}
\end{eqnarray}
Then, we can single a closed differential equation of $c_{4}(t)$ out from the system of first-order differential equations as
\begin{equation}
i\dot{c}_{4}(t)
=
e^{i(\nu+\Delta\nu)t}\Omega,
\quad
c_{4}(0)=0,
\end{equation}
and we obtain
\begin{equation}
c_{4}(t)
=
\frac{\Omega}{\nu+\Delta\nu}
[1-e^{i(\nu+\Delta\nu)t}].
\end{equation}
Moreover, we delete terms that include $\Omega^{2}/(\nu+\Delta\nu)$ from the differential equation of $\dot{c}_{0}(t)$
because of $\Omega^{2}/(\nu+\Delta\nu)\ll 1$.

At this moment, we obtain the following system of first-order differential equations:
\begin{eqnarray}
i\dot{c}_{0}(t)
&=&
-i e^{i\Delta\nu t}\eta\Omega c_{3}(t), \nonumber \\
i\dot{c}_{3}(t)
&=&
e^{-i(\nu+\Delta\nu)t}\Omega[2x(t)-i\eta e^{i\nu t}(1-2c_{0}(t))], \nonumber \\
i\dot{x}(t)
&=&
e^{i(\nu+\Delta\nu)t}\Omega c_{3}(t),
\end{eqnarray}
\begin{equation}
c_{0}(0)=1,
\quad
c_{3}(0)=0,
\quad
x(0)=0.
\end{equation}
Further, making replacements,
\begin{eqnarray}
\tilde{c}_{3}(t)
&=&
e^{i\Delta\nu t}c_{3}(t), \nonumber \\
\tilde{x}(t)
&=&
e^{-i\nu t}x(t),
\end{eqnarray}
we finally arrive at the following system of first-order differential equations:
\begin{equation}
\dot{\mbox{\boldmath $c$}}(t)=A\mbox{\boldmath $c$}(t)+\mbox{\boldmath $b$},
\end{equation}
\begin{equation}
\mbox{\boldmath $c$}(t)
=
\left(
\begin{array}{c}
c_{0}(t) \\
\tilde{x}(t) \\
\tilde{c}_{3}(t)
\end{array}
\right),
\quad
A
=
\left(
\begin{array}{ccc}
0 & 0 & -\eta\Omega \\
0 & -i\nu & -i\Omega \\
2\eta\Omega & -2i\Omega & i\Delta\nu \\
\end{array}
\right),
\quad
\mbox{\boldmath $b$}
=
\left(
\begin{array}{c}
0 \\
0 \\
-\eta\Omega
\end{array}
\right).
\end{equation}

We express eigenvalues of the matrix $A$ as $\lambda_{1}$, $\lambda_{2}$, and $\lambda_{3}$.
The functions $c_{0}(t)$, $\tilde{x}(t)$, and $\tilde{c}_{3}(t)$ are superpositions of $\exp(\lambda_{i}t)$ for $i=1,2,3$.
More specifically, we can write down $c_{0}(t)$, $\tilde{x}(t)$, and $\tilde{c}_{3}(t)$ in the form,
\begin{equation}
a_{1}\exp(\lambda_{1}t)+a_{2}\exp(\lambda_{2}t)+a_{3}\exp(\lambda_{3}t).
\label{c0-x-c3-superposition-0}
\end{equation}
The eigenvalues $\lambda_{i}$ for $i=1,2,3$ depend on $A$ but not on $\mbox{\boldmath $b$}$.
The coefficients $a_{i}$ for $i=1,2,3$ given in Eq.~(\ref{c0-x-c3-superposition-0}) depend on $\mbox{\boldmath $b$}$.
Thus, the periods of $c_{0}(t)$, $\tilde{x}(t)$, and $\tilde{c}_{3}(t)$, that is, the period of $|\psi(t)\rangle$, are given by
$\left|
\mbox{Re}
\left[
2\pi i/\lambda_{k}
\right]
\right|$
for $k=1,2,3$.
We describe the longest period among them as ${\cal T}_{0}(\Omega,\eta,\nu,\Delta\nu)$ and it is given by
\begin{eqnarray}
{\cal T}_{0}(\Omega,\eta,\nu,\Delta\nu)
&=&
12 \pi\mbox{Im}[1/X], \nonumber \\
X
&=&
2i(\Delta\nu-\nu)+\frac{2^{4/3}Y}{Z}-2^{2/3}Z, \nonumber \\
Z
&=&
(iU+\sqrt{W})^{1/3}, \nonumber \\
W
&=&
4Y^{3}-U^{2}, \nonumber \\
Y
&=&
\Delta\nu^{2}+\nu\Delta\nu+\nu^{2}+6(1+\eta^{2})\Omega^{2}, \nonumber \\
U
&=&
2\Delta\nu^{3}+3\nu\Delta\nu^{2}-3\nu^{2}\Delta\nu-2\nu^{3} \nonumber \\
&&
+
18(\Delta\nu+\eta^{2}\Delta\nu-\nu+2\eta^{2}\nu)\Omega^{2}.
\label{period-T-0-formula-0}
\end{eqnarray}

According to Fig.~\ref{Figure1},
letting $\Omega=0.1$, $\eta=0.025$, $\Delta\nu=0.025$, and $\nu=5$,
we obtain ${\cal T}_{0}=14790.7$ and it matches well with the numerical result obtained in Fig.~\ref{Figure1}.
Hence, we can conclude that ${\cal T}_{0}(\Omega,\eta,\nu,\Delta\nu)$ represents a mathematical form of a good approximate period of $|\psi(t)\rangle$
whose inital state is given by $|g\rangle_{\mbox{\scriptsize A}}|g\rangle_{\mbox{\scriptsize B}}|0\rangle_{\mbox{\scriptsize P}}$.

\begin{figure}
\begin{center}
\includegraphics{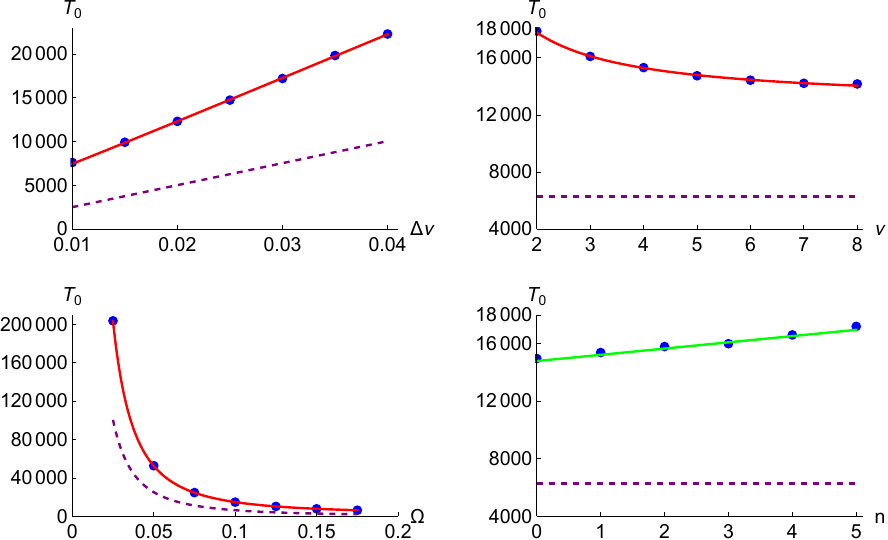}
\end{center}
\caption{
Plots of the periods of the time evolution of the quantum gates,
$T_{0}(n)$, ${\cal T}_{0}(\Omega,\eta,\nu,\Delta\nu)$, ${\cal T}'(n,\Omega,\eta,\nu,\Delta\nu)$, and $T_{\mbox{\scriptsize MS}}$,
where ${\cal T}'(n,\Omega,\eta,\nu,\Delta\nu)$ is given by \ref{appendix-initial-non-zero-phonon-states}.
Blue points represent $T_{0}(n)$ obtained by solving the Schr{\"{o}}dinger equation for the second-order perturbation theory with the fourth-order Runge--Kutta method
as explained in Sec.~\ref{section-simplified-MS-gate}
where the initial state is given by $|\psi(t=0)\rangle=|g\rangle_{\mbox{\scriptsize A}}|g\rangle_{\mbox{\scriptsize B}}|n\rangle_{\mbox{\scriptsize P}}$.
The solid red curves represent ${\cal T}_{0}(\Omega,\eta,\nu,\Delta\nu)$ given by Eq.~(\ref{period-T-0-formula-0}).
The solid green line represents Eq.~(\ref{T0-n-dependence-line-approximation-0}).
The dashed purple curve represents $T_{\mbox{\scriptsize MS}}$ for the MS gate.
(a)
$T_{0}$(0), ${\cal T}_{0}$, and $T_{\mbox{\scriptsize MS}}$ as functions of $\Delta\nu$ for $n=0$, $\Omega=0.1$, $\nu=5$, and $\eta=0.025$.
(b)
$T_{0}$(0), ${\cal T}_{0}$, and $T_{\mbox{\scriptsize MS}}$ as functions of $\nu$ for $n=0$, $\Omega=0.1$, $\Delta\nu=0.025$, and $\eta=0.025$.
(c)
$T_{0}(0)$, ${\cal T}_{0}$, and $T_{\mbox{\scriptsize MS}}$ as functions of $\Omega$ for $n=0$, $\nu=5$, $\Delta\nu=0.025$, and $\eta=0.025$.
(d)
$T_{0}(n)$, ${\cal T}'(n,\Omega,\eta,\nu,\Delta\nu)$, and $T_{\mbox{\scriptsize MS}}$ as functions of $n$ for $\Delta\nu=0.025$, $\Omega=0.1$, $\nu=5$, and $\eta=0.025$.
}
\label{Figure3}
\end{figure}

In Fig.~\ref{Figure3}, we compare ${\cal T}_{0}$ given by Eq.~(\ref{period-T-0-formula-0}) with $T_{0}(n)$ obtained numerically
by solving the Schr{\"{o}}dinger equation for the second-order perturbation theory with the fourth-order Runge--Kutta method
in Sec.~\ref{section-simplified-MS-gate}.
Figure~\ref{Figure3}(a) shows plots of ${\cal T}_{0}$ and $T_{0}(0)$ as functions of $\Delta\nu$ for $n=0$, $\Omega=0.1$, $\eta=0.025$, and $\nu=5$.
The solid red curve represent ${\cal T}_{0}$ given by Eq.~(\ref{period-T-0-formula-0}) and blue points represent $T_{0}(0)$.
Both of them match each other well.
The dashed purple curve represents $T_{\mbox{\scriptsize MS}}$ given by Eq.~(\ref{definition-T-MS-0}).
Looking at these plots, we note that the period of our proposed gate is almost twice as long as $T_{\mbox{\scriptsize MS}}$.
Figure~\ref{Figure3}(b) shows plots of ${\cal T}_{0}$ and $T_{0}(0)$ as functions of $\nu$ for $n=0$, $\Omega=0.1$, $\eta=0.025$, and $\Delta\nu=0.025$.
Figure~\ref{Figure3}(c) shows plots of ${\cal T}_{0}$ and $T_{0}(0)$ as functions of $\Omega$ for $n=0$, $\eta=0.025$, $\Delta\nu=0.025$, and $\nu=5$.
In Figs.~\ref{Figure3}(b) and (c), ${\cal T}_{0}$ and $T_{0}(0)$ match well with each other.

In \ref{appendix-initial-non-zero-phonon-states}, we estimate the period of the time evolution of the proposed gate whose initial state is given by
$|\psi(t=0)\rangle=|g\rangle_{\mbox{\scriptsize A}}|g\rangle_{\mbox{\scriptsize B}}|n\rangle_{\mbox{\scriptsize P}}$
for $n=1,2,3,...$.
We describe them as ${\cal T}'(n,\Omega,\eta,\nu,\Delta\nu)$.
Letting
$\Omega=0.1$, $\eta=0.025$, $\nu=5$, and $\Delta\nu=0.025$,
we compute ${\cal T}'(n)$ for $n=0,1,...,5$ and plot them as a function of $n$ with the solid green curve in Fig.~\ref{Figure3}(d).
This curve is approximated by a function of $n$ as
\begin{equation}
{\cal T}'(n)\simeq14790+433 n.
\label{T0-n-dependence-line-approximation-0}
\end{equation}
Thus, we can judge that the dependence of ${\cal T}'(n)$ on $n$ is small.
The solid green curve and the blue points match well with each other.

\section{\label{section-example-two-qubit-gate}An example of a two-qubit gate that generates entanglement but is strongly vulnerable to the thermal fluctuations of the phonons}
It is not difficult to implement a two-qubit gate that generates entanglement between ions A and B.
However, if we try to make the two-qubit gate robust under the thermal fluctuations of the phonons,
we notice that it is very difficult.
This section gives a concrete example of a two-qubit gate that is strongly sensitive to the thermal noise of the phonons.

Let us consider a system that consists of ions A, B, and phonons.
We assume that its Hamiltonian is given by
\begin{eqnarray}
H
&=&
H_{\mbox{\scriptsize A}}+\kappa H_{\mbox{\scriptsize B}}, \nonumber \\
H_{\mbox{\scriptsize A}}
&=&
i\alpha(\sigma_{\mbox{\scriptsize A}+}a-\sigma_{\mbox{\scriptsize A}-}a^{\dagger}), \nonumber \\
H_{\mbox{\scriptsize B}}
&=&
i\alpha(\sigma_{\mbox{\scriptsize B}+}a^{\dagger}-\sigma_{\mbox{\scriptsize B}-}a).
\label{Hamiltonian-toy-model-0}
\end{eqnarray}
where $\alpha=\eta\Omega$
and $\kappa$ is a dimensionless parameter.
We can obtain $H_{\mbox{\scriptsize A}}$ and $H_{\mbox{\scriptsize B}}$ defined by Eq.~(\ref{Hamiltonian-toy-model-0})
by applying the rotating-wave approximation to Eq.~(\ref{ion-trap-Hamiltonian-basic-0}).
For the sake of simplicity, we assume $\kappa>0$.
Here, we prepare the initial state of the system given by
$|g\rangle_{\mbox{\scriptsize A}}|g\rangle_{\mbox{\scriptsize B}}|0\rangle_{\mbox{\scriptsize P}}$,
so that we only need to consider three orthonormal vectors
$\{
|g\rangle_{\mbox{\scriptsize A}}|g\rangle_{\mbox{\scriptsize B}}|0\rangle_{\mbox{\scriptsize P}},
|e\rangle_{\mbox{\scriptsize A}}|e\rangle_{\mbox{\scriptsize B}}|0\rangle_{\mbox{\scriptsize P}},
|g\rangle_{\mbox{\scriptsize A}}|e\rangle_{\mbox{\scriptsize B}}|1\rangle_{\mbox{\scriptsize P}}
\}$
for examining the time evolution of $|\psi(t)\rangle$.
We describe these three vectors as
\begin{eqnarray}
|g\rangle_{\mbox{\scriptsize A}}|g\rangle_{\mbox{\scriptsize B}}|0\rangle_{\mbox{\scriptsize P}}
&=&
(1,0,0)^{\mbox{\scriptsize T}}, \nonumber \\
|e\rangle_{\mbox{\scriptsize A}}|e\rangle_{\mbox{\scriptsize B}}|0\rangle_{\mbox{\scriptsize P}}
&=&
(0,1,0)^{\mbox{\scriptsize T}}, \nonumber \\
|g\rangle_{\mbox{\scriptsize A}}|e\rangle_{\mbox{\scriptsize B}}|1\rangle_{\mbox{\scriptsize P}}
&=&
(0,0,1)^{\mbox{\scriptsize T}}.
\end{eqnarray}
Then, the Hamiltonian is written down in the form of the following $3\times 3$ matrix:
\begin{equation}
H
=
i\alpha
\left(
\begin{array}{ccc}
0 & 0 & -\kappa \\
0 & 0 & 1 \\
\kappa & -1 & 0 \\
\end{array}
\right).
\end{equation}

\begin{figure}
\begin{center}
\includegraphics[width=0.5\linewidth]{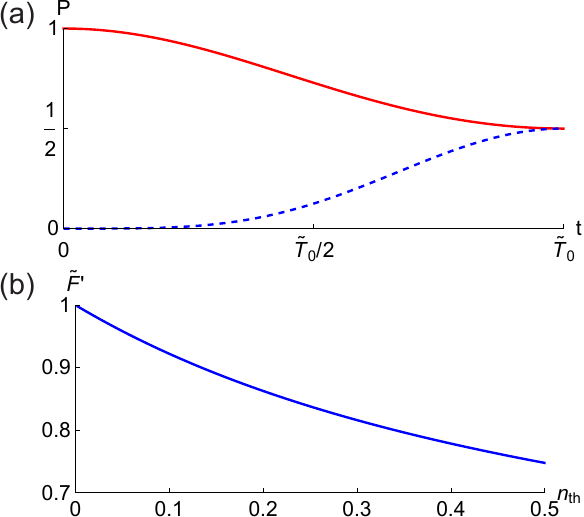}
\end{center}
\caption{(a) Plots of $P_{gg0}=|c_{0}(t)|^{2}$ and $P_{ee0}=|c_{1}(t)|^{2}$ as functions of $t$
for $\alpha=1$ and $\kappa=\sqrt{2}-1$.
The solid red and dashed blue curves represent $P_{gg0}$ and $P_{ee0}$, respectively.
We can confirm that $P_{gg0}=P_{ee0}=1/2$ at $t=\tilde{T}_{0}$.
(b) A plot of $\tilde{F}'$ as a function of $n_{\mbox{\scriptsize th}}$ for $\kappa=\sqrt{2}-1$ and $t=\tilde{T}_{0}$.
The shape of the graph does not depend on $\alpha$.}
\label{Figure4}
\end{figure}

We can compute the time evolution of
$|\psi(t)\rangle$ for $t\geq 0$ as follows:
\begin{equation}
|\psi(t)\rangle
=
c_{0}(t)|g\rangle_{\mbox{\scriptsize A}}|g\rangle_{\mbox{\scriptsize B}}|0\rangle_{\mbox{\scriptsize P}}
+
c_{1}(t)|e\rangle_{\mbox{\scriptsize A}}|e\rangle_{\mbox{\scriptsize B}}|0\rangle_{\mbox{\scriptsize P}}
+
c_{2}(t)|g\rangle_{\mbox{\scriptsize A}}|e\rangle_{\mbox{\scriptsize B}}|1\rangle_{\mbox{\scriptsize P}},
\label{time-evolution-1}
\end{equation}
\begin{eqnarray}
c_{0}(t)
&=&
[1+\kappa^{2}\cos(\alpha\sqrt{1+\kappa^{2}}t)]/(1+\kappa^{2}), \nonumber \\
c_{1}(t)
&=&
\kappa[1-\cos(\alpha\sqrt{1+\kappa^{2}}t)]/(1+\kappa^{2}), \nonumber \\
c_{2}(t)
&=&
\kappa\sin(\alpha\sqrt{1+\kappa^{2}}t)/\sqrt{\kappa^{2}+1}.
\label{time-evolution-2}
\end{eqnarray}
Assuming $c_{0}(t_{0})=1/\sqrt{2}$,
we obtain $t_{0}$ as
\begin{equation}
t_{0}
=
\frac{1}{\alpha\sqrt{1+\kappa^{2}}}\arccos \theta_{0},
\label{definition-t0}
\end{equation}
\begin{equation}
\theta_{0}
=
\frac{-2+\sqrt{2}+\sqrt{2}\kappa^{2}}{2\kappa^{2}}.
\end{equation}
Assuming $c_{1}(t_{1})=1/\sqrt{2}$,
we obtain $t_{1}$ as
\begin{equation}
t_{1}
=
\frac{1}{\alpha\sqrt{1+\kappa^{2}}}\arccos \theta_{1},
\end{equation}
\begin{equation}
\theta_{1}
=
\frac{-\sqrt{2}+2\kappa-\sqrt{2}\kappa^{2}}{2\kappa}.
\end{equation}
Setting $t_{0}=t_{1}$, we obtain $\kappa=\sqrt{2}-1$.
Then defining $\tilde{T}_{0}=t_{0}=t_{1}$ with $\kappa=\sqrt{2}-1$, we obtain
\begin{equation}
\tilde{T}_{0}
=
\frac{\sqrt{2+\sqrt{2}}\pi}{2\alpha}.
\label{T0-definition}
\end{equation}
The period of $|\psi(t)\rangle$ is given by $\tilde{T}_{\mbox{\scriptsize period}}=4\tilde{T}_{0}$.
Figure~\ref{Figure4}(a) shows graphs of $P_{gg0}=|c_{0}(t)|^{2}$ and $P_{ee0}=|c_{1}(t)|^{2}$ as functions of $t$
for $\alpha=1$ and $\kappa=\sqrt{2}-1$.

We define a fidelity of the state that evolves from
$|\psi(0)\rangle=|g\rangle_{\mbox{\scriptsize A}}|g\rangle_{\mbox{\scriptsize B}}|0\rangle_{\mbox{\scriptsize P}}$
and one of the Bell states at $t=\tilde{T}_{0}$ as
\begin{equation}
\tilde{F}
=
|{}_{\mbox{\scriptsize AB}}\langle\Phi^{+}|{}_{\mbox{\scriptsize P}}\langle 0|\psi(\tilde{T}_{0})\rangle|^{2},
\end{equation}
\begin{equation}
|\Phi^{+}\rangle_{\mbox{\scriptsize AB}}
=
(1/\sqrt{2})(
|g\rangle_{\mbox{\scriptsize A}}|g\rangle_{\mbox{\scriptsize B}}
+
|e\rangle_{\mbox{\scriptsize A}}|e\rangle_{\mbox{\scriptsize B}}
).
\end{equation}
From Eqs.~(\ref{time-evolution-1}) and (\ref{time-evolution-2}), we obtain
$\tilde{F}=1$ for $\kappa=\sqrt{2}-1$.
Figure~\ref{Figure4}(b) shows a graph of the fidelity for the state with warm ions
as a function of the average number of thermal phonons $n_{\mbox{\scriptsize th}}$
for $\kappa=\sqrt{2}-1$ and $t=\tilde{T}_{0}$.
Strictly speaking, here, we have redefined the fidelity $\tilde{F}'$ as
\begin{equation}
\tilde{F}'
=
\sum_{n=0}^{\infty}
p_{n}
|{}_{\mbox{\scriptsize AB}}\langle\Phi^{+}|{}_{\mbox{\scriptsize P}}\langle n|\psi_{n}(t)\rangle|^{2},
\end{equation}
where
$|\psi_{n}(t)\rangle$ denotes the state of the ions A, B and the phonons whose initial state is given by
$|g\rangle_{\mbox{\scriptsize A}}|g\rangle_{\mbox{\scriptsize B}}|n\rangle_{\mbox{\scriptsize P}}$.

We draw attention that the shape of the graph in Fig.~\ref{Figure4}(b) does not depend on $\alpha$.
Looking at Fig.~\ref{Figure4}(b),
we note that the curve of $\tilde{F}'$ has a sharp peak at $n_{\mbox{\scriptsize th}}=0$,
that is to say, the graph is convex downward.
This characteristic is a complete contrast to that of the curves in Fig.~\ref{Figure2}.
Thus, we can conclude that the two-qubit gate caused by the Hamiltonian defined by
Eq.~(\ref{Hamiltonian-toy-model-0})
is not well protected from thermal fluctuations of the phonons.

\section{\label{section-conclusions-discussions}Conclusions and discussions}
In this paper, we propose the simplified MS gate that is caused by applying the two monochromatic coherent light beams to the two ions A and B individually and simultaneously.
The length of the execution time of this proposed gate for generating the maximally entangled state is comparable to that of the original MS gate.
(Precisely, it is twice as long as $T_{\mbox{\scriptsize MS}}$.)
Thus, we can estimate the length of the execution time of the proposed gate approximately at
$2T_{\mbox{\scriptsize MS}}
=
\pi\Delta\nu/(\eta^{2}\Omega^{2})
$
because of Eq.~(\ref{definition-T-MS-0}).
On the other hand, the execution time of Cirac and Zoller's ion--phonon quantum gate is given by
$T_{\mbox{\scriptsize CZ2}}\gg 1/(\eta^{2}\nu)$,
where $\nu=\omega_{\mbox{\scriptsize eg}}-\omega$,
$\hbar\omega_{\mbox{\scriptsize eg}}$ represents the difference of energy levels between excited and ground states of the ion,
and $\omega$ represents the angular frequency of a laser beam applied to the ion.
The derivation of this relation is given in \ref{appendix-operation-times-Cirac-Zoller-gate}.
Because $T_{\mbox{\scriptsize MS}}$ is on the order of $\eta^{-2}$,
we can consider that $T_{\mbox{\scriptsize MS}}$ is much shorter than $T_{\mbox{\scriptsize CZ2}}$.

As explained in Sec.~\ref{section-example-two-qubit-gate}, in general,
it is difficult to construct a two-qubit gate that generates the maximally entangled state and is protected against fluctuations in the number of phonons.
Drawing the curve of the fidelity as a function of the average number of the thermal phonons for our proposed gate in Fig.~\ref{Figure2},
we cannot show that the derivative of the curve is equal to zero.
Thus, we must recognize that our method is not insensitive to thermal noise of the phonons.
Hence, we reach the following conclusions.
The original MS gate is insensitive to thermal fluctuation,
our proposed simplified MS gate is weakly sensitive,
and the hypothetical gate shown in Sec.~\ref{section-example-two-qubit-gate} is strongly sensitive.

In Sec.~\ref{section-mathematical-analyses} and \ref{appendix-initial-non-zero-phonon-states},
to estimate the periods of the time evolution of the system given rise to by the proposed gate,
we utilize results obtained by numerical simulations empirically.
This is because solving the Schr{\"{o}}dinger equation with the Hamiltonian given by Eq.~(\ref{Hamiltonian-AB-second-order-1})
as the second-order perturbation theory is very difficult.
The proposed gate cannot reduce the execution time for generating the maximally entangled state compared to the original MS gate.
This is one of our future challenges.

An application of our simplified MS gate is as follows.
We consider a factory where we can produce many imperfect Bell pairs of ions with our proposed gate.
Performing entanglement purification, we obtain pure Bell states with high fidelity.
For these tasks, our gate can be useful because purification operations remove errors caused by thermal noise of phonons.
In this way, if we combine our gate with error correction protocols, it can be a practical tool for generation of entanglement.

At the present moment, the trapped ion quantum computer is one of the promising candidates for NISQ processing.
Many methods for quantum computation are great, however, what makes the trapped ion quantum computer different is the robustness
of the two-qubit gate against thermal fluctuations.
For example, a noisy thermal environment is fatal for quantum computation with the Josephson junctions.
In \cite{Jing2022}, experimental demonstrations of gate-controlled quantum dots based on two-dimensional materials were discussed.
According to \cite{Jing2022}, experiments of quantum dots for quantum computation also require the temperature of the system to be kept low enough.

\appendix

\section{\label{appendix-time-evolution-initial-states}Time evolution for the initial states
$|e\rangle_{\mbox{\scriptsize A}}|e\rangle_{\mbox{\scriptsize B}}|n\rangle_{\mbox{\scriptsize P}}$,
$|g\rangle_{\mbox{\scriptsize A}}|e\rangle_{\mbox{\scriptsize B}}|n\rangle_{\mbox{\scriptsize P}}$,
and
$|e\rangle_{\mbox{\scriptsize A}}|g\rangle_{\mbox{\scriptsize B}}|n\rangle_{\mbox{\scriptsize P}}$
for $n=0,1,...$}
We can derive the time evolution of the state of the proposed gate for
$\Delta\nu>0$ and $\Omega>0$
from results of numerical calculations empirically as follows.
If
$|\psi(t=0)\rangle
=
|e\rangle_{\mbox{\scriptsize A}}|e\rangle_{\mbox{\scriptsize B}}|n\rangle_{\mbox{\scriptsize P}}$ for $n=0,1,...$,
we obtain
\begin{eqnarray}
|\psi(t)\rangle
&\simeq&
\frac{1}{2}
[1-\exp(-2\pi i\frac{t}{T_{0}(n)})]
|g\rangle_{\mbox{\scriptsize A}}|g\rangle_{\mbox{\scriptsize B}}|n\rangle_{\mbox{\scriptsize P}} \nonumber \\
&&
+
\frac{1}{2}
[1+\exp(-2\pi i\frac{t}{T_{0}(n)})]
|e\rangle_{\mbox{\scriptsize A}}|e\rangle_{\mbox{\scriptsize B}}|n\rangle_{\mbox{\scriptsize P}}.
\end{eqnarray}
If
$|\psi(t=0)\rangle
=
|g\rangle_{\mbox{\scriptsize A}}|e\rangle_{\mbox{\scriptsize B}}|n\rangle_{\mbox{\scriptsize P}}$
for $n=0,1,...$,
we obtain
\begin{equation}
|\psi(t)\rangle
\simeq
\exp(2\pi i\frac{t}{T_{ge}})
|g\rangle_{\mbox{\scriptsize A}}|e\rangle_{\mbox{\scriptsize B}}|n\rangle_{\mbox{\scriptsize P}},
\end{equation}
where
$T_{ge}=T_{ge}(n,\Omega,\nu,\Delta\nu,\eta)$,
and $T_{ge}$ is on the order of $\nu/\Omega^{2}$.
For example, for $n=0$, $\Omega=0.1$, $\nu=5$, $\Delta\nu=0.025$, and $\eta=0.025$,
we obtain $T_{ge}=1575$.
In general, $T_{ge}\ll T_{0}$ holds.
Moreover, $T_{ge}$ decreases as $n$ increases.
When $n=0$ holds, $T_{ge}$ does not depend on $\eta$.
We cannot estimate a specific mathematical expression of $T_{ge}$
from results of numerical calculations.
As an exception, for $n=0$, we obtain
\begin{equation}
T_{ge}
\simeq
\frac{\pi}{\Omega^{2}}(\nu+\Delta\nu).
\end{equation}
If
$|\psi(t=0)\rangle
=
|e\rangle_{\mbox{\scriptsize A}}|g\rangle_{\mbox{\scriptsize B}}|n\rangle_{\mbox{\scriptsize P}}$
for $n=0,1,...$,
we obtain
\begin{equation}
|\psi(t)\rangle
\simeq
\exp(-2\pi i\frac{t}{T_{eg}})
|e\rangle_{\mbox{\scriptsize A}}|g\rangle_{\mbox{\scriptsize B}}|n\rangle_{\mbox{\scriptsize P}},
\end{equation}
where
$T_{eg}=T_{eg}(n,\Omega,\nu,\Delta\nu,\eta)$,
and $T_{eg}$ is on the order of $\nu/\Omega^{2}$
and
$T_{eg}\ll T_{0}$ holds.
Further, $T_{eg}$ decreases as $n$ increases.
For example, for $n=0$, $\Omega=0.1$, $\nu=5$, $\eta=0.025$, and $\Delta\nu=0.025$,
$T_{eg}$ is given by $1442$.
We cannot estimate a specific mathematical expression of $T_{eg}$ from the results of the numerical calculations.

\section{\label{appendix-first-order-differential-equation}An explicit form of the Schr{\"{o}}dinger equation derived from the second-order perturbation theory}
An explicit form of the Schr{\"{o}}dinger equation
derived from Eq.~(\ref{Hamiltonian-AB-second-order-1})
is given as follows:
\begin{eqnarray}
i\dot{c}_{0}(t)/\Omega
&=&
-(1/2)(\eta^{2}-2) e^{i (\nu+\Delta\nu)t}c_{2}(t)
-i \eta e^{i \Delta\nu t}c_{3}(t) \nonumber \\
&&
-(1/2)(\eta^{2}-2) e^{-i (\nu+\Delta\nu)t}c_{4}(t)
-i \eta e^{-i (2\nu+\Delta\nu)t}c_{5}(t), \nonumber \\
i\dot{c}_{1}(t)/\Omega
&=&
-i \eta 
e^{i (2\nu+\Delta\nu)t}c_{2}(t)
-(1/2) (\eta^{2}-2)
e^{i (\nu+\Delta\nu)t}c_{3}(t) \nonumber \\
&&
-i \eta e^{-i \Delta\nu t}c_{4}(t)
-(1/2) (\eta^{2}-2) e^{-i (\nu+\Delta\nu)t}c_{5}(t), \nonumber \\
i\dot{c}_{2}(t)/\Omega
&=&
-(1/2) (\eta^{2}-2) 
e^{-i (\nu+\Delta\nu)t}c_{0}(t)
+i \eta e^{-i (2\nu+\Delta\nu)t}c_{1}(t) \nonumber \\
&&
-(1/2) (\eta^{2}-2) e^{-i 
(\nu+\Delta\nu)t}c_{6}(t)
-i \eta e^{-i (2\nu+\Delta\nu)t}c_{7}(t), \nonumber \\
i\dot{c}_{3}(t)/\Omega
&=&
i 
\eta e^{-i \Delta\nu t}c_{0}(t)
-(1/2) (\eta^{2}-2) e^{-i (\nu+\Delta\nu)t}c_{1}(t) \nonumber \\
&&
-i \eta 
e^{-i \Delta\nu t}c_{6}(t)
-(1/2) (\eta^{2}-2) 
e^{-i (\nu+\Delta\nu)t}c_{7}(t), \nonumber \\
i\dot{c}_{4}(t)/\Omega
&=&
-(1/2) (\eta^{2}-2)
e^{i (\nu+\Delta\nu)t}c_{0}(t)
+i \eta e^{i \Delta\nu 
t}c_{1}(t) \nonumber \\
&&
-(1/2) (\eta^{2}-2) e^{i 
(\nu+\Delta\nu)t}c_{6}(t)
-i \eta e^{i \Delta\nu t}c_{7}(t), \nonumber \\
i\dot{c}_{5}(t)/\Omega
&=&
i \eta 
e^{i (2\nu+\Delta\nu)t}c_{0}(t)
-(1/2) (\eta^{2}-2) e^{i (\nu+\Delta\nu)t}c_{1}(t) \nonumber \\
&&
-i \eta 
e^{i (2\nu+\Delta\nu)t}c_{6}(t)
-(1/2) (\eta^{2}-2) 
e^{i (\nu+\Delta\nu)t}c_{7}(t), \nonumber \\
i\dot{c}_{6}(t)/\Omega
&=&
-(1/2) (\eta^{2}-2) e^{i (\nu+\Delta\nu)t}c_{2}(t)
+i \eta 
e^{i \Delta\nu t}c_{3}(t) \nonumber \\
&&
-(1/2) (\eta^{2}-2) 
e^{-i (\nu+\Delta\nu)t}c_{4}(t)
+i \eta e^{-i (2\nu+\Delta\nu)t}c_{5}(t), \nonumber \\
i\dot{c}_{7}(t)/\Omega
&=&
i \eta e^{i (2\nu+\Delta\nu)t}c_{2}(t)
-(1/2) 
(\eta^{2}-2) e^{i (\nu+\Delta\nu)t}c_{3}(t) \nonumber \\
&&
+i \eta 
e^{-i \Delta\nu t}c_{4}(t)
-(1/2) (\eta^{2}-2)
e^{-i (\nu+\Delta\nu)t}c_{5}(t).
\end{eqnarray}

\section{\label{appendix-initial-non-zero-phonon-states}Estimation of the periods of the proposed gate
with the initial states
$|g\rangle_{\mbox{\scriptsize A}}|g\rangle_{\mbox{\scriptsize B}}|n\rangle_{\mbox{\scriptsize P}}$ for $n=1, 2, ...$}
We estimate the period of the time evolution of the proposed gate whose initial state is given by
$|\psi(t=0)\rangle=|g\rangle_{\mbox{\scriptsize A}}|g\rangle_{\mbox{\scriptsize B}}|n\rangle_{\mbox{\scriptsize P}}$
for $n=1,2,3,...$.
For the sake of simplicity, we assume that the system lies in a twelve-dimensional Hilbert space whose orthonormal basis is given by
\begin{eqnarray}
&&
|g\rangle_{\mbox{\scriptsize A}}|g\rangle_{\mbox{\scriptsize B}}|k\rangle_{\mbox{\scriptsize P}},
\quad
|g\rangle_{\mbox{\scriptsize A}}|e\rangle_{\mbox{\scriptsize B}}|k\rangle_{\mbox{\scriptsize P}}, \nonumber \\
&&
|e\rangle_{\mbox{\scriptsize A}}|g\rangle_{\mbox{\scriptsize B}}|k\rangle_{\mbox{\scriptsize P}},
\quad
|e\rangle_{\mbox{\scriptsize A}}|e\rangle_{\mbox{\scriptsize B}}|k\rangle_{\mbox{\scriptsize P}}
\quad
\mbox{for $k=n-1,n,n+1$}.
\end{eqnarray}
We describe these states as $|\bar{i}\rangle$ for $i=0,1,...,11$ in order,
that is,
$|\bar{0}\rangle=|g\rangle_{\mbox{\scriptsize A}}|g\rangle_{\mbox{\scriptsize B}}|n-1\rangle_{\mbox{\scriptsize P}}$,
$|\bar{1}\rangle=|g\rangle_{\mbox{\scriptsize A}}|g\rangle_{\mbox{\scriptsize B}}|n\rangle_{\mbox{\scriptsize P}}$,
...,
$|\bar{11}\rangle=|e\rangle_{\mbox{\scriptsize A}}|e\rangle_{\mbox{\scriptsize B}}|n+1\rangle_{\mbox{\scriptsize P}}$.
The annihilation operator of the phonons is approximated by the $3\times 3$ matrix,
\begin{equation}
\hat{a}
=
\left(
\begin{array}{ccc}
0 & \sqrt{n} & 0 \\
0 & 0 & \sqrt{n+1} \\
0 & 0 & 0 \\
\end{array}
\right),
\end{equation}
where we set the basis as $\{|n-1\rangle_{\mbox{\scriptsize P}},|n\rangle_{\mbox{\scriptsize P}},|n+1\rangle_{\mbox{\scriptsize P}}\}$.
We compute the period approximately in a similar way to the previous case where the initial state is given by
$|g\rangle_{\mbox{\scriptsize A}}|g\rangle_{\mbox{\scriptsize B}}|0\rangle_{\mbox{\scriptsize P}}$
in Sec.~\ref{section-mathematical-analyses}.
We write down $|\psi(t)\rangle$ as a superposition of $|\bar{i}\rangle$ for $i=0,1,...,11$,
\begin{equation}
|\psi(t)\rangle
=
\sum_{i=0}^{11}c_{i}(t)|\bar{i}\rangle.
\end{equation}
Next, from Eq.~(\ref{Hamiltonian-AB-second-order-1}),
we derive matrix elements
$\langle\bar{j}|\hat{H}|\bar{k}\rangle$
($j,k\in\{0,1,...,11\}$)
up to the second-order perturbation of $\eta$.
From the Schr{\"{o}}dinger equation, we can derive the system of first-order differential equations.

First, in the system of first-order differential equations, we replace terms
$-2+n\eta^{2}$, $-2+(1+n)\eta^{2}$, $-2+(1+2n)\eta^{2}$ with $-2$.
Second, we put an approximate relationship between
the coefficient $c_{1}(t)$
of $|g\rangle_{\mbox{\scriptsize A}}|g\rangle_{\mbox{\scriptsize B}}|n\rangle_{\mbox{\scriptsize P}}$
and
the coefficient $c_{10}(t)$
of $|e\rangle_{\mbox{\scriptsize A}}|e\rangle_{\mbox{\scriptsize B}}|n\rangle_{\mbox{\scriptsize P}}$
as
\begin{equation}
c_{10}(t)\simeq 1-c_{1}(t).
\end{equation}
Third, for the coefficient $c_{3}(t)$
of
$|g\rangle_{\mbox{\scriptsize A}}|e\rangle_{\mbox{\scriptsize B}}|n-1\rangle_{\mbox{\scriptsize P}}$
and the coefficient $c_{8}(t)$ of
\\
$|e\rangle_{\mbox{\scriptsize A}}|g\rangle_{\mbox{\scriptsize B}}|n+1\rangle_{\mbox{\scriptsize P}}$,
we set the following approximations:
\begin{equation}
c_{3}(t)\simeq 0,
\quad
c_{8}(t)\simeq 0.
\end{equation}
Fourth, according to the rotating-wave approximation,
we delete terms that include $e^{\pm 2i\nu t}$.
Fifth, we delete terms that include $\eta^{2}\Omega$.
Sixth, for the coefficient $c_{0}(t)$
of $|g\rangle_{\mbox{\scriptsize A}}|g\rangle_{\mbox{\scriptsize B}}|n-1\rangle_{\mbox{\scriptsize P}}$
and the coefficient $c_{9}(t)$ of
$|e\rangle_{\mbox{\scriptsize A}}|e\rangle_{\mbox{\scriptsize B}}|n-1\rangle_{\mbox{\scriptsize P}}$,
we put approximate relations,
\begin{eqnarray}
x(t)
&=&
\frac{1}{2}[c_{0}(t)+c_{9}(t)], \nonumber \\
c_{0}(t)-c_{9}(t)&\simeq &0.
\end{eqnarray}
Seventh, at this moment, we can single a closed differential equation of $c_{4}(t)$
of
$|g\rangle_{\mbox{\scriptsize A}}|e\rangle_{\mbox{\scriptsize B}}|n\rangle_{\mbox{\scriptsize P}}$
out from the system of first-order differential equations derived from the Schr{\"{o}}dinger equation as
\begin{equation}
i\dot{c}_{4}(t)
=
e^{-i(\nu+\Delta\nu)t}\Omega,
\quad
c_{4}(0)=0,
\end{equation}
and we obtain
\begin{equation}
c_{4}(t)
=
\frac{\Omega}{\nu+\Delta\nu}
[-1+e^{-i(\nu+\Delta\nu)t}].
\end{equation}
Eighth, for the coefficient $c_{2}(t)$
of
$|g\rangle_{\mbox{\scriptsize A}}|g\rangle_{\mbox{\scriptsize B}}|n+1\rangle_{\mbox{\scriptsize P}}$
and the coefficient $c_{11}(t)$ of
$|e\rangle_{\mbox{\scriptsize A}}|e\rangle_{\mbox{\scriptsize B}}|n+1\rangle_{\mbox{\scriptsize P}}$,
we put approximate relations,
\begin{eqnarray}
y(t)
&=&
\frac{1}{2}[c_{2}(t)+c_{11}(t)], \nonumber \\
c_{2}(t)-c_{11}(t)&\simeq &0.
\end{eqnarray}
Ninth, at this moment, we can single a closed differential equation of $c_{7}(t)$
of
$|e\rangle_{\mbox{\scriptsize A}}|g\rangle_{\mbox{\scriptsize B}}|n\rangle_{\mbox{\scriptsize P}}$
out from the system of differential equations as
\begin{equation}
i\dot{c}_{7}(t)
=
e^{i(\nu+\Delta\nu)t}\Omega,
\quad
c_{7}(0)=0,
\end{equation}
and we obtain
\begin{equation}
c_{7}(t)
=
\frac{\Omega}{\nu+\Delta\nu}
[1-e^{i(\nu+\Delta\nu)t}].
\end{equation}
Tenth, we delete terms that include $\Omega^{2}/(\nu+\Delta\nu)$ from the differential equation of $c_{1}(t)$ of
$|g\rangle_{\mbox{\scriptsize A}}|g\rangle_{\mbox{\scriptsize B}}|n\rangle_{\mbox{\scriptsize P}}$.
Finally, we replace
the coefficient
$c_{5}(t)$ of
$|g\rangle_{\mbox{\scriptsize A}}|e\rangle_{\mbox{\scriptsize B}}|n+1\rangle_{\mbox{\scriptsize P}}$,
the coefficient
$c_{6}(t)$ of
$|e\rangle_{\mbox{\scriptsize A}}|g\rangle_{\mbox{\scriptsize B}}|n-1\rangle_{\mbox{\scriptsize P}}$,
$x(t)$, and $y(t)$ with
\begin{eqnarray}
\tilde{c}_{5}(t)&=&e^{i\Delta\nu t}c_{5}(t), \nonumber \\
\tilde{c}_{6}(t)&=&e^{-i\Delta\nu t}c_{6}(t), \nonumber \\
\tilde{y}(t)&=&e^{-i\nu t}y(t), \nonumber \\
\tilde{x}(t)&=&e^{i\nu t}x(t).
\end{eqnarray}

Now, we arrive at the following system of first-order differential equations:
\begin{equation}
\dot{\mbox{\boldmath $c$}}(t)=A\mbox{\boldmath $c$}(t)+\mbox{\boldmath $b$},
\end{equation}
\begin{eqnarray}
&&
\mbox{\boldmath $c$}(t)
=
\left(
\begin{array}{c}
\tilde{x}(t) \\
c_{1}(t) \\
\tilde{y}(t) \\
\tilde{c}_{5}(t) \\
\tilde{c}_{6}(t) \\
\end{array}
\right),
\quad
\mbox{\boldmath $b$}
=
\left(
\begin{array}{c}
0 \\
0 \\
0 \\
-i\Omega\sqrt{n+1}\eta \\
-i\Omega\sqrt{n}\eta \\
\end{array}
\right), \nonumber \\
&&
A
=
\left(
\begin{array}{ccccc}
i\nu & 0 & 0 & 0 &-2i\Omega \\
0 & 0 & 0 & 2\sqrt{n+1}\eta\Omega & 2\sqrt{n}\eta\Omega \\
0 & 0 & -i\nu & -2i\Omega & 0 \\
0 & -\sqrt{n+1}\eta\Omega & -i\Omega & i\Delta & 0 \\
-i\Omega & -\sqrt{n}\eta\Omega & 0 & 0 & -i\Delta \\
\end{array}
\right).
\end{eqnarray}
Because $A$ is a $5\times 5$ matrix,
it is very hard to obtain its eigenvalues as closed-form solutions.
Thus, we compute the eigenvalues of the matrix $A$ numerically.
Expressing the eigenvalues of $A$ as $\lambda_{k}$ for $k=1,...,5$,
we calculate periods as
$\left|
\mbox{Re}
\left[
2\pi i/\lambda_{k}
\right]
\right|$
for $k=1,...,5$.
We write the longest one among these periods
as ${\cal T}'(n,\Omega,\eta,\nu,\Delta\nu)$.

\section{\label{appendix-operation-times-Cirac-Zoller-gate}Estimation of the operation time for Cirac and Zoller's ion--phonon gate}
In this section, we evaluate the operation times of Cirac and Zoller's one-ion gate and ion--phonon gate.
Analyses of this section are based on investigations given by \cite{Jonathan2000}.

\begin{figure}
\begin{center}
\includegraphics{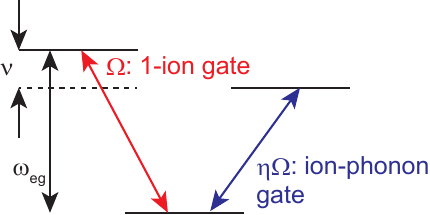}
\end{center}
\caption{
A schematic of energy levels for Cirac and Zoller's one-qubit and ion--phonon gates.
}
\label{Figure5}
\end{figure}

We apply laser light with an angular frequency $\omega=\omega_{\mbox{\scriptsize eg}}$
and the Rabi frequency $\Omega$ to the ion for carrying out the one-ion gate.
Thus, its coupling constant appearing in the Hamiltonian is given by $\Omega$.
Contrastingly, to perform the ion--phonon quantum gate,
we let laser light with an angular frequency $\omega=\omega_{\mbox{\scriptsize eg}}-\nu$
and the Rabi frequency $\Omega$ illuminates the ion.
Thus, its coupling constant appearing in the Hamiltonian is given by $\eta\Omega$.
To execute the one-ion and ion--phonon gates stably,
the two processes shown in Fig.~\ref{Figure5} must be distinguishable from each other.
Thus, we obtain a condition
\begin{equation}
\Omega\eta\ll\nu.
\label{gate-operation-time-condition-1}
\end{equation}
Hence, describing operation times of Cirac and Zoller's one-qubit and ion--phonon gates
as $T_{\mbox{\scriptsize CZ}1}$ and $T_{\mbox{\scriptsize CZ}2}$, respectively,
we obtain
\begin{equation}
T_{\mbox{\scriptsize CZ}1}\sim\Omega^{-1},
\quad
T_{\mbox{\scriptsize CZ}2}\sim(\eta\Omega)^{-1},
\quad
T_{\mbox{\scriptsize CZ}2}\gg T_{\mbox{\scriptsize CZ}1},
\label{CiracZoller-operation-times-0}
\end{equation}
\begin{equation}
T_{\mbox{\scriptsize CZ}2}\gg\nu^{-1}.
\label{CiracZoller-operation-times-1}
\end{equation}
Equation~(\ref{CiracZoller-operation-times-1}) implies that the operation time for generating entanglement
with Cirac and Zoller's ion--phonon quantum gate is much longer than $\nu^{-1}$.

We have obtained Eq.~(\ref{gate-operation-time-condition-1}) as the condition of Cirac and Zoller's ion--phonon quantum gate.
However, this condition becomes useless as $\eta\to 0$.
To overcome this problem,
we return to the original Hamiltonian of Cirac and Zoller's ion--phonon quantum gate.
If we apply a laser field with $\delta=-\nu$,
that is to say, with a frequency $\omega=\omega_{\mbox{\scriptsize eg}}-\nu$ to the ion,
the Hamiltonian defined in Eq.~(\ref{ion-trap-Hamiltonian-basic-0}) is rewritten in the form under $\eta\to 0$
\begin{equation}
H'
=
\hbar\Omega
[
\sigma_{+}\exp(i\nu t)
+
\sigma_{-}\exp(-i\nu t)
].
\end{equation}

Assuming that the initial state of the ion A and the phonon mode P is given by $|g\rangle_{\mbox{\scriptsize A}}|0\rangle_{\mbox{\scriptsize P}}$,
we evaluate the correction of the energy with the perturbation theory.
We obtain the first-order energy shift as
\begin{equation}
{}_{\mbox{\scriptsize A}}\langle g|
{}_{\mbox{\scriptsize P}}\langle 0|
H'
|g\rangle_{\mbox{\scriptsize A}}|0\rangle_{\mbox{\scriptsize P}}
=
0.
\end{equation}
The second-order energy shift is given by
\begin{equation}
-\sum_{m\neq 0}
\frac{|\langle\psi_{m}|H'|\psi_{0}\rangle|^{2}}{E_{m}-E_{0}},
\label{second-order-energy-shift-CiracZoller-0} 
\end{equation}
where $|\psi_{0}\rangle$ denotes the initial state,
$E_{0}$ represents the energy level of $|\psi_{0}\rangle$,
$|\psi_{m}\rangle$ denotes the intermediate state,
and $E_{m}$ represents the energy level of $|\psi_{m}\rangle$.
Substitutions of the initial state $|\psi_{0}\rangle=|g\rangle_{\mbox{\scriptsize A}}|0\rangle_{\mbox{\scriptsize P}}$,
the energy of the laser field $E_{0}=\hbar(\omega_{\mbox{\scriptsize eg}}-\nu)$,
the intermediate state $|\psi_{m}\rangle=|e\rangle_{\mbox{\scriptsize A}}|0\rangle_{\mbox{\scriptsize P}}$,
and the excitation energy of the ion $E_{m}=\hbar\omega_{\mbox{\scriptsize eg}}$
into Eq.~(\ref{second-order-energy-shift-CiracZoller-0} ) gives the second-order energy shift in the form
\begin{equation}
-\hbar
\frac{|\Omega \exp(i\nu t)|^{2}}{\nu}
=
-\frac{\hbar\Omega^{2}}{\nu}.
\end{equation}

To ensure that the quantum gate works correctly,
we must let the second-order correction of the energy be much smaller
than the product of the coupling constant $\eta\Omega$ and $\hbar$.
Thus, we obtain another condition
\begin{equation}
\Omega\ll \eta\nu.
\label{gate-operation-time-condition-2}
\end{equation}
This condition is stricter than that of Eq.~(\ref{gate-operation-time-condition-1}).
Thus, we attain
\begin{equation}
T_{\mbox{\scriptsize CZ}2}\gg\frac{1}{\eta^{2}\nu}.
\label{Cirac-Zoller-two-qubit-gate-operation-time}
\end{equation}

\section*{Acknowledgements}
This work was supported by MEXT Quantum Leap Flagship Program (MEXT Q-LEAP) Grant Number JPMXS0120351339.

\section*{Data availability statement}
The data obtained by numerical calculations and C++ programs will be available from the author upon reasonable request.

\section*{ORCID iDs}
Hiroo Azuma https://orcid.org/0000-0002-4374-8727
\\

\end{document}